\documentclass[aps, prl, twocolumn,10pt, superscriptaddress, groupedaddress,showpacs]{revtex4-1}

\usepackage{amsmath}
\usepackage{amssymb}
\usepackage{graphicx}
\usepackage[usenames,dvipsnames]{xcolor}
\usepackage[colorlinks, linkcolor = blue, citecolor = blue, filecolor = blue, urlcolor = blue, bookmarks=true, breaklinks=true]{hyperref}

\renewcommand{\vec}[1]{\mathbf{#1}} 

\begin{document}


\title{Disentangling intra-cycle interferences in photoelectron momentum distributions using orthogonal two-colour laser fields}

\author{Xinhua\,Xie$^1$}
\author{Tian\,Wang$^2$}
\author{ShaoGang\,Yu$^3$}
\author{XuanYang\,Lai$^3$}
\author{Stefan\,Roither$^1$}
\author{Daniil\,Kartashov$^1$}
\author{Andrius\,Baltu\v{s}ka$^1$}
\author{XiaoJun\,Liu$^3$}
\email{xjliu@wipm.ac.cn}
\author{Andr\'e\,Staudte$^2$}
\email{andre.staudte@nrc-cnrc.gc.ca}
\author{Markus\,Kitzler$^1$}
\email{markus.kitzler@tuwien.ac.at}

\affiliation{$^1$Photonics Institute, Technische Universit\"at Wien, 1040 Vienna, Austria}
\affiliation{$^2$Joint Laboratory for Attosecond Science of the National Research Council and the University of Ottawa, Ottawa, Ontario, Canada K1A 0R6}
\affiliation{$^3$State Key Laboratory of Magnetic Resonance and Atomic and Molecular Physics, Wuhan Institute of Physics and Mathematics, Chinese Academy of Sciences, Wuhan 430071, China}

\begin{abstract}
We use orthogonally polarized two-colour (OTC) laser pulses to separate quantum paths in multiphoton ionization of Ar atoms. Our OTC pulses consist of 400~nm and 800~nm light at a relative intensity ratio of 10:1. We find a hitherto unobserved interference in the photoelectron momentum distribution, which exhibits a strong dependence on the relative phase of the OTC pulse. Analysis of model calculations reveal that the interference is caused by quantum pathways from non-adjacent quarter cycles.
\end{abstract}


\maketitle

Key processes in nature such as the breakage and formation of molecular bonds are ultimately determined by the dynamics of the molecular valence electrons. 
Obtaining access to the dynamics of the valence electrons in atoms and molecules thus constitutes a crucial capability for understanding and manipulating molecular processes. 
An ideal tool to interrogate and drive the rearrangement of the valence electron cloud in such systems on its intrinsic time-scale of attoseconds is the strong electric field of intense laser pulses.
The momentum distributions of photoelectrons emitted during the interaction of such fields with atoms and molecules contain a wealth of information about the structure and field-induced attosecond dynamics in the target.

Here, we use tailored laser fields, so-called orthogonally polarized two-color (OTC) fields \cite{Kitzler2005, Kitzler2007, Kitzler2008NJP}, that allow determining the direction of the emission of photoelectron wavepackets on laser sub-cycle time-scales. 
This allows us to gain control over the creation of  interference structures in photoelectron momentum distributions and at the same time permits to experimentally disentangle a novel class of interference fringes from the many other structures in the measured distributions.
%
%
These structures are, on the one hand, due to the spatial properties of the ionizing system, e.g., the symmetry of the ionizing orbital, that modulate the emitted photoelectron wavepacket directly during the ionization step and are imprinted in the final photoelectron momentum distributions \cite{Comtois2013, Meckel2014}.
On the other hand, the observed photoelectron momentum distributions are also determined by the field-driven dynamics of the emitted wavepackets during which their amplitude and phase structure is modified, thus, determining  their interference at the detector.

The observed photoelectron momentum distribution therefore  constitutes a complicated interference pattern of all emitted wavepackets and contains information about the dynamics of the bound states from which they are coherently split by ionization, their field-driven dynamics, and their interaction with the parent ion's combined nuclear and electronic charge distribution. 
The dynamical and structural information contained in the photoelectron momentum distributions can often 
be extracted by analysing them in terms of semiclassical electron trajectories \cite{Milosevic2006}. 
Within this framework, the measured distributions can be understood as interferences of electron wavepackets that reach the detector on direct or recolliding trajectories, the latter scattered either in the forward or backward direction, giving rise to inter-cycle and intra-cycle fringes \cite{Xie2012_interferometry, Arbo2010}, holographic structures \cite{Huismans2011Science, Haertelt2016PRL} and diffraction patterns \cite{Ray2008PRL, Okunishi2008PRL, Meckel2008}, respectively.

While the diffraction patterns generated by the recolliding trajectories can be exploited for obtaining structural information about the molecule with few femtosecond resolution \cite{Xu2010b, Blaga2012Nature, Pullen2015NatComm, Wolter2016}, it has been shown that measured interference fringes due to the direct trajectories can be inverted to yield the dynamics of valence electron wavepackets with few attosecond resolution \cite{Xie2012_interferometry}.
Two types of 
direct 
trajectories need to be distinguished for such inversion: those launched with a periodicity of exactly one laser cycle giving rise to inter-cycle fringes also referred to as  above-threshold ionization (ATI) peaks \cite{Agostini1979, Freeman1991}, and trajectories that originate from a single laser cycle, leading to intra-cycle interference fringes \cite{Xie2012_interferometry, Arbo2010}.
The resulting modulation of the inter-cycle peaks by the intra-cycle fringes can be interpreted as the diffraction of the emitted wavepackets on a temporal grating \cite{Arbo2010}. 
Thus far only intra-cycle fringes that originate from electron wavepackets emitted during adjacent quarter-cycles of the optical wave
--- a situation that has been called the temporal double slit --- have been considered theoretically \cite{Arbo2006a, Arbo2010, Arbo2012} and observed in experiments  \cite{Xie2012_interferometry, Lindner2005PRL, Gopal2009}.


In our experiment OTC fields 
 are used for the generation of a customized temporal grating for wavepacket emission, which allows us to create and observe for the first time intra-cycle interferences from non-adjacent quarter-cycles. 
The key to this achievement is the generation of an OTC field that consists of a weak orthogonal component with half the frequency of the ionizing field. 
By varying the phase-delay between the two colours this enables us to control the momentum overlap of photoelectron wavepackets originating from adjacent and non-adjacent quarter-cycles.
Depending on the phase-delay, the two-dimensional character of the OTC fields maps the sub-cycle wavepacket emission times into different regions of the momentum space, 
which solves the notorious problem of disentangling the different types of interference structures from measured  photoelectron momentum distributions and allows the clear observation of the intra-cycle fringes from non-adjacent quarter-cycles.


Disentanglement of different quantum paths using the two-dimensional mapping provided by OTC fields has been considered previously \cite{Zhang2014PRA, Richter2015PRL, Geng2015PRL, Richter2016PRA}. Previously employed OTC pulses were generated by temporally overlapping an intense 800~nm pulse with a weaker, perpendicularly polarized, 400~nm pulse. In contrast, we invert the intensity ratio between long and short wave length and use an intensity ratio of approximately 10:1 for 400~nm and 800~nm, respectively. Whereas multiphoton ionization in OTC pulses with a half-frequency streaking field has been investigated previously theoretically \cite{Li2015PRA}, our work explores this concept in  experiment and theory to find a previously unidentified sub-cycle interference structure in the photoelectron momentum  distribution.

\begin{figure}[tb]
\centering
\includegraphics[width=0.95\columnwidth, angle=0]{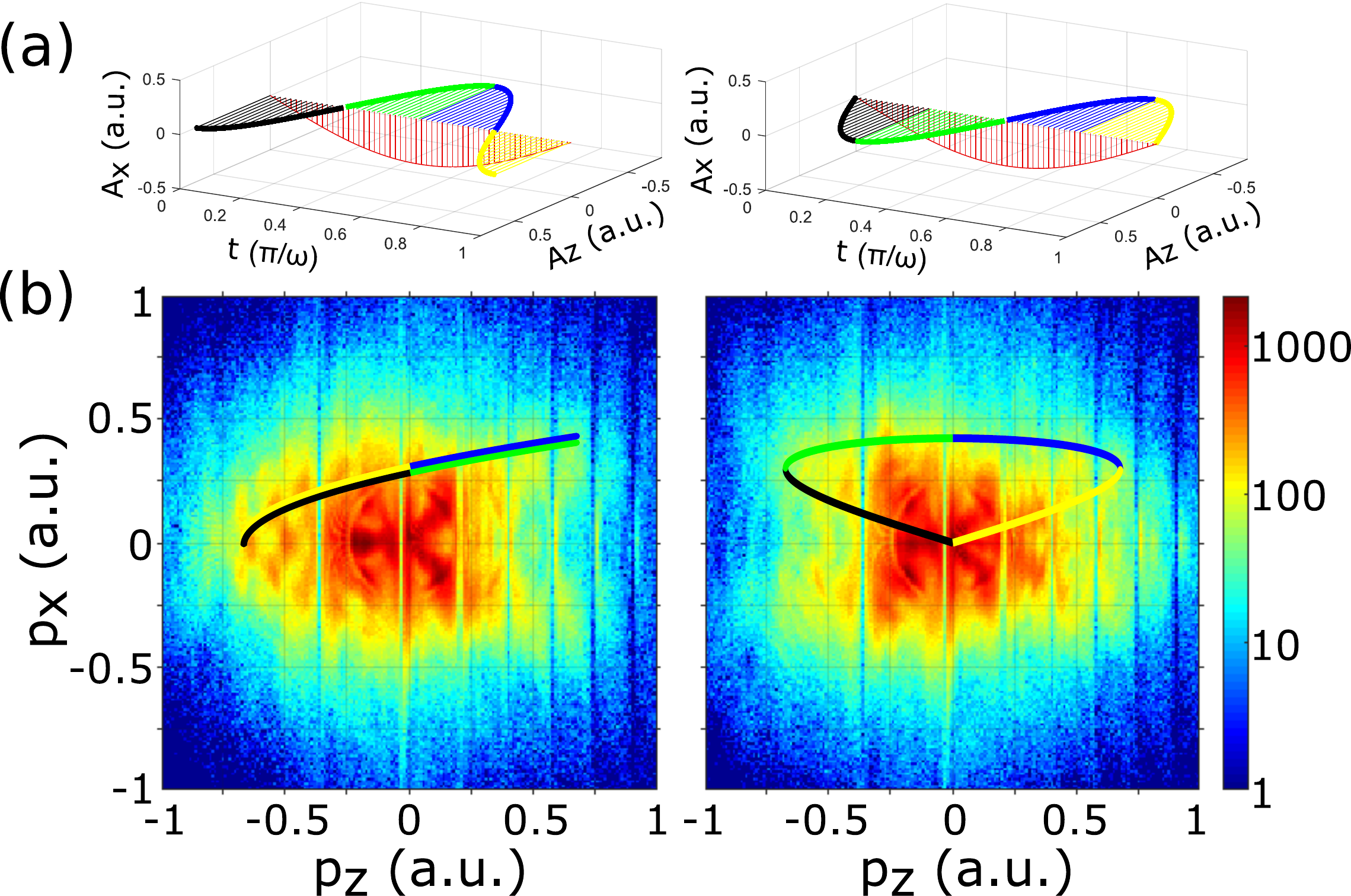}
\caption{(a) Vector potential of an OTC field for two phase delays $\varphi=1.5\pi$ (left) and $\varphi=\pi$ (right). Shown is one optical cycle of the 400~nm field, each quarter cycle is color-coded. (b) Measured photoelectron momentum distributions of Ar ionized in a 45~fs OTC pulse, for the phase delays in (a). Superimposed on the momentum distributions is the SMM prediction of the electron momentum,  $\vec{p} = -\vec{A}(t_i)$, for $\vec{A}(t)$ from panel (a).}\label{fig1}
\end{figure}

In the experiment, OTC pulses were generated by combining s-polarized, 45~fs, 800~nm laser pulses, with p-polarized, 45~fs, 400~nm pulses, in a collinear geometry at a repetition rate of 5~kHz. The peak intensity in the isolated 800\,nm and 400\,nm pulses was estimated (uncertainty about 20\%) as $0.2 \times 10^{14} \mathrm{W/cm}^2$ and $2 \times 10^{14} \mathrm{W/cm}^2$, respectively. We used COLTRIMS \cite{Doerner2000PhysRep} to record the three-dimensional (3D) photoelectron momentum distribution of single ionization of argon.
As usual, we define the field of the OTC pulse as \cite{Zhang2014PRA}
\begin{equation}
\vec{E}(t)=\hat{E}
\left[
f_{x}(t)\cos(\omega t)\vec{e}_x+f_{z}(t)\cos(2\omega t+ \varphi)\vec{e}_z
\right]\label{OTC_field}
\end{equation}
with $\hat{E}$ the peak field strength, $\omega$ the laser frequency of the 800~nm field, $f_{x,z}$ the pulse envelopes along the 800~nm and 400~nm direction, respectively, and $\varphi$ the relative phase between the two colors.
Fig.~\ref{fig1} illustrates how the orthogonally superposed 800~nm streaks the sub-cycle ionization in the 400~nm field.
According to the simple man's model (SMM)  \cite{VanLindenvandenHeuvell1988, Corkum1989} the final electron momentum after the laser pulse is given by $\vec{p} = -\vec{A}(t_i)$, with $\vec{A}(t)=-\int_{-\infty}^t \vec{E}(t') \text{d}t'$ the laser field's vector potential and $t_i$ the birth time of the electron.
Fig.~\ref{fig1}(a) depicts $\vec{A}(t)$ for each wavelength separately and for the two important limiting cases $ \varphi = 1.5 \pi$ and $\pi$. 
Fig.~\ref{fig1}(b) shows the corresponding measured photoelectron momentum distributions in the polarization plane of the OTC pulse. 
The minima in the momentum distributions for $p_{z,400nm} = -0.375, -0.05, 0.2, 0.4, 0.55, 0.75, 0.9$~a.u. are due to the confining magnetic field in the spectrometer and therefore are an experimental artifact.
On top of the photoelectron momentum distributions we superimposed $\vec{p} = -\vec{A}(t_i)$ for $t_i$ covering half of an optical cycle of the 800~nm field using the color code of Fig.~\ref{fig1}(a).

\noindent

\begin{figure}[tb]
\centering
\includegraphics[width=0.95\columnwidth, angle=0]{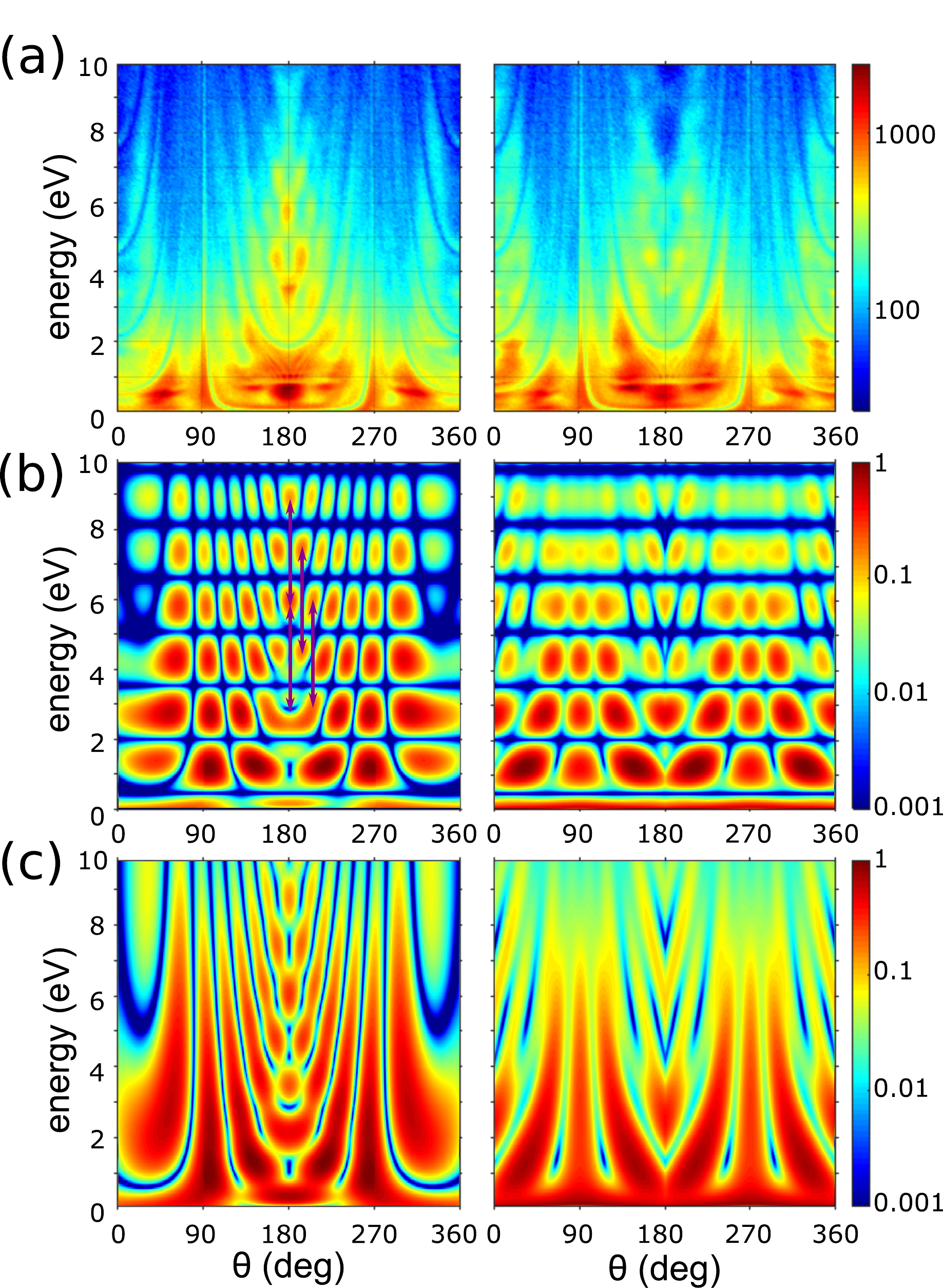}
\caption{Angle-resolved energy spectra of photoelectrons for a phase delay $\varphi=1.5\pi$ (left) and $\varphi=\pi$ (right). (a) Experiment, (b) SFA calculation using Eq.~(\ref{eq_Mp}), (c) SFA calculation with the OTC pulse restricted to one optical cycle of 800~nm field. The arrows indicate peaks separated by 3~eV. See text for details.}\label{fig2}
\end{figure}

In order to facilitate the analysis of the complex momentum distributions it is useful to transform the data into a polar coordinate system in the plane of polarization with the photoelectron energy $E_{kin}$ as the radial coordinate. In Fig.~\ref{fig2}(a) we have done this transformation for the data of Fig.~\ref{fig1}(b). A polar angle of $\theta=0^{\circ}$ corresponds to a momentum towards the right side, i.e., $p_{z,400nm}>0$ and $p_{x,800nm}=0$~a.u. Apart from the sharp line-features due to the experimental artifacts mentioned above, both spectra show pronounced structures and differ most significantly at $\theta=180^{\circ}$. Whereas for a phase of $\varphi = \pi$ the photoelectron spectrum is only faintly structured, a rich interference carpet appears for a phase of $\varphi = 1.5 \pi$. In the latter, a series of energy peaks with a spacing of about 3~eV are observed at $\theta=180^{\circ}, 180^{\circ}\pm 15^{\circ}$, and $180^{\circ}\pm 25^{\circ}$.

To understand these interference structures, we turn to the strong-field approximation (SFA) theory. In the SFA theory, the transition amplitude of the photoelectron from the initial bound state $\left| {{\psi_0}} \right\rangle$ to the final Volkov state $\left|{\psi_\textbf{p}^\text{V}}\right\rangle$ with the final momentum $\textbf{p}$ is given by  \cite{Reiss1980PRA, Milosevic1998PRA}
\begin{equation}
M_\textbf{p}=-i\int_{-\infty}^{\infty}\text{d}t
\langle\textbf{p}+\textbf{A}(t)
\left|\textbf{r}\cdot\textbf{E}(t)\right| \psi_0(r)\rangle e^{iS(\textbf{p},t)} ,
\label{SFA_trans_prob}
\end{equation}
where $\textbf{r}\cdot\textbf{E}(t)$ is the laser-field-electron interaction,
$S(\textbf{p},t)=-\frac{1}{2}\int_{t}^\infty dt'[\textbf{p}+\textbf{A}(t')]^2+I_pt$ is the semiclassical action, and $I_p$ denotes the ionization potential. For sufficiently high intensity and low frequency of the laser pulse, the temporal integration in Eq.~(\ref{SFA_trans_prob}) can be evaluated with high accuracy by the saddle-point method \cite{Figueira2008PRA}. The corresponding transition amplitude then becomes
\begin{equation}
M_\textbf{p}\sim\sum\limits_{s}A_{s}(\textbf{p}) e^{iS(\textbf{p},t_s)} \quad,
\label{eq_Mp}
\end{equation}
where  $A_{s}(\textbf{p})=\sqrt{\frac{2 \pi i}{\partial^2 S_\textbf{p}/\partial  t_s^2}} \left \langle\textbf{p}+\textbf{A}(t_s) \left|\textbf{r}\cdot\textbf{E}(t_s)\right| \psi_0(r)\right \rangle $, and the index $s$ runs over the relevant saddle points obtained by solving the saddle point equation $[\textbf{p}+\textbf{A}(t)]^2=-2I_p$. Each saddle point corresponds to a quantum orbit. 
Physically, the transition amplitude $M_\textbf{p}$ represents the coherent superposition of all quantum orbits  \cite{Becker2002AdvAtMolOptPhys, Figueira2002PRA}. 
Eq.~(\ref{eq_Mp}) thus provides an intuitive formulation of   interference patterns in photoelectron spectra.

\noindent

In Fig.~\ref{fig2}(b) we show the photoelectron spectrum $\frac{d|M_\textbf{p}|^2}{dE_{kin}d\theta}$ calculated using  Eq.~(\ref{eq_Mp}) for the two phase delays $\varphi=1.5\pi$ and $\pi$ with the laser parameters of the experiment (peak intensities for 800/400~nm are 0.2/$2\times 10^{14}$~W/cm$^2$) and the $I_p$ of argon decreased by 1.07~eV (to 14.69~eV) for obtaining agreement with the peak energies observed in the experiment \cite{Paulus2001a}. Our calculations exhibit very distinct energy bands with a spacing of 1.5~eV, which are due to an inter-cycle interference that arises from the 800~nm component of the OTC field. However, for $\varphi=1.5\pi$ the energy interval along $\theta=180^{\circ}$ is about 3~eV. At $\theta=180^{\circ}\pm13^{\circ}$ a second set of peaks is present with an energy interval of 3~eV and an offset of about 1.5~eV.
Furthermore, a third set of peaks, also with an energy interval of 3~eV, appears at about $\theta=180^{\circ}\pm24^{\circ}$. There is no energy offset with respect to the first set of peaks at $\theta=180^{\circ}$. These distinct features, marked by arrows in Fig.~\ref{fig2}(b), are in qualitative agreement with the experimental observations in Fig.~\ref{fig2}(a). 
Below we will show that the appearance of the peaks with the energy interval of 3~eV at $\theta=180^{\circ}$ is due to the inter-cycle interference of the 400~nm field, while the second and third set of peaks are closely related to a newly uncovered intra-cycle interference, i.e., a non-adjacent intra-cycle interference, the focus of this Letter.

The temporal grating created by the 800~nm component dominates the calculated energy spectrum. This discrepancy is due to the fact that the simulation in Fig.~\ref{fig2}(b) assumes a continuous-wave (cw) OTC field, whereas the experiment has a pulse envelope with a full width at half maximum of intensity of 45~fs. 
In order to remove the cw effect in the simulation, we follow a procedure that was used previously for a linearly polarized two-color field \cite{Xie2012_interferometry} and restrict the ionization time of quantum orbits in Eq.~(\ref{eq_Mp}) to one cycle of the 800~nm field for the remainder of the discussion. 
The resulting photoelectron spectrum is shown in Fig.~\ref{fig2}(c). Thereby, a better agreement with the experimental data is achieved. In comparison to Fig.~\ref{fig2}(b), it is clear that the inter-cycle interference features from the 800~nm field, i.e., the ATI structures with energy spacing of 1.5~eV, disappear. On the other hand, the prominent interference carpet feature at around $180^{\circ}$, as observed in the experiment, still persists. 
A closer inspection reveals some discrepancy between the experimental data and the SFA simulation, for example in the relative electron yields for different angle and energy ranges.
These can be ascribed to the influence of the parent ion's Coulomb potential, which is neglected in the present SFA theory but in OTC fields 
can have a significant influence on the relative electron yields emitted along different directions \cite{Yu2016a}.

\begin{figure}[tb]
\centering
\includegraphics[width=0.95\columnwidth, angle=0]{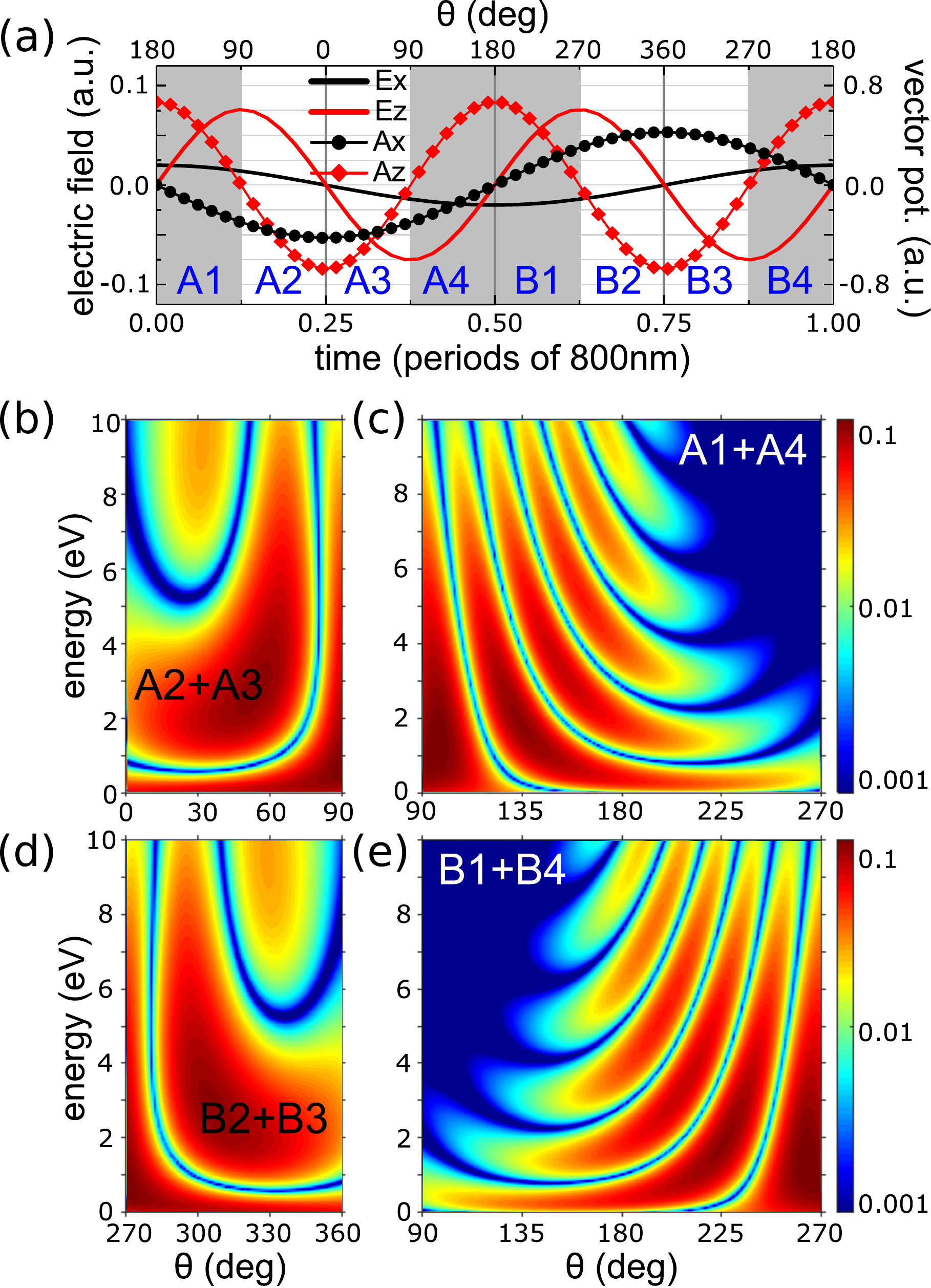}
\caption{(a) Electric fields ($E_x$, $E_z$) and vector potentials ($A_x$, $A_z$) for $\varphi=1.5\pi$  within one cycle of 800~nm. The characters A1-A4 and B1-B4 denote the different ionization time windows. The top axis shows the asymptotic emission angle $\theta$ of the photoelectron according to the simple man's model $\vec{p} = -\vec{A}(t)$.  (b-e) SFA calculations using Eq.~(\ref{eq_Mp}) with the quantum orbits from the different ionization time windows as indicated in the figures.}\label{fig3}
\end{figure}

To understand the origin of the interference carpet at around $180^{\circ}$ observed for $\varphi=1.5\pi$, we divide the electric field and vector potential of the one-cycle OTC pulse into 8 parts, i.e., A1-A4 and B1-B4, as shown in Fig.~\ref{fig3}(a). The top axis in Fig.~\ref{fig3}(a) indicates the limits of the emission angle $\theta$ in the polarization plane for each quarter cycle of the 400~nm component of the OTC pulse based on the mapping of the ionization time to final momentum within the SMM, i.e., $\vec{p} = -\vec{A}(t)$. 
For example, the electron emitted from A1 and A4 with $A_z>0$ and $A_x<0$ will arrive at the detector with a final momentum $p_z<0$ and $p_x>0$, contributing to the spectra at an angle of $90^{\circ}<\theta<180^{\circ}$. 
In Fig.~\ref{fig3}(b) our SFA analysis confirms that photoelectrons from within the ionization time windows A2 and A3 of our OTC pulse will be detected at an angle of $0^{\circ}<\theta<90^{\circ}$. However, photoelectrons emitted within the ionization time windows A1 and A4 are detected at $90^{\circ}<\theta<270^{\circ}$, 
well outside their classically permitted region [see Fig.~\ref{fig3}(c)]. Similar results can be also found for other regions of spectra shown in Figs.~\ref{fig3}(d) and (e).

Comparing with the left panel of Fig.~\ref{fig2}(c), the interference patterns at $0^{\circ}<\theta<90^{\circ}$ are due to interfering electron trajectories emitted from adjacent quarter cycles of the 400~nm (A2 and A3). In contrast, the stripes at $90^{\circ}<\theta<270^{\circ}$ visible for $\varphi=1.5\pi$ in Fig.~\ref{fig2}(c) 
are from the interference of two orbits from non-adjacent quarters of the 400~nm field (A1 and A4, B1 and B4). The first kind of intra-cycle interference has been widely investigated in the literature \cite{Arbo2010, Xie2012_interferometry, Lai2013PRA, Li2015SR}. In contrast, the second type (i.e., non-adjacent) intra-cycle interference has not been  observed or identified before, to the best of our knowledge.
%

Our SFA simulations show that the interference fringes from this novel non-adjacent intra-cycle interference are much finer than that of the adjacent intra-cycle interference. The reason is that the ionization delays 
of the electron orbits for the non-adjacent intra-cycle interference are larger than those of the adjacent ones and, thus, the change of the corresponding action difference of the two orbits is more sensitive to the momentum of the photoelectron. This result is analogous to the double-slit interference \cite{Lindner2005PRL}, for which the stripe width on the screen is inversely proportional to the distance of the two slits. It is this novel intra-cycle interference, through which the fine and bent fringes at $90^{\circ}<\theta<270^{\circ}$ for $\varphi=1.5\pi$ are formed. Together with the inter-cycle interference by the 800~nm field, they give rise to the appearance of the second and third sets of peaks at $167^{\circ}$ and $156^{\circ}$, respectively, with an energy interval of 3~eV and an energy offset of 1.5~eV with respect to each other [see Figs.~\ref{fig2}(a) and (c)].

\noindent

It is worth noting that the above-discussed non-adjacent intra-cycle interference is strongly dependent on the phase delay of the OTC pulse. For example, the right panels of Figs.~\ref{fig2}(a) and (b) show that for the OTC pulse with $\varphi = \pi$, the pronounced interference structure disappears in the experiment and the corresponding SFA simulation. The reason is that for the laser pulse with $\varphi = \pi$, only the traditional adjacent intra-cycle interference from the adjacent quarters 
can exist. Therefore, after removing the inter-cycle interference in Fig.~\ref{fig2}(c) by restricting the  integration to one cycle, no clear carpet structure can be observed in the spectrum for an OTC pulse with $\varphi = \pi$.

One may argue that this novel intra-cycle interference from the non-adjacent quarters should also exist and might be observed for a single-color field (see also recent work by Maxwell~et~al. \cite{Maxwell2017}). However, for a single-color field, the two kinds of the intra-cycle interference structures usually appear in the same regions of the momentum space. Thus, the overlap of the two types of intra-cycle interferences obscures the non-adjacent intra-cycle interference in the spectra. In contrast, by adding a weak 800\,nm field with $\varphi = 1.5 \pi$ in the $x$ direction, as we have demonstrated in this Letter, the two orbits corresponding to the adjacent intra-cycle interference will be driven to different quarters of the photoelectron spectra and thus the non-adjacent intra-cycle interference can be easily disentangled and be observed in the experiments.

Concluding, we have demonstrated that the temporal grating in an OTC pulse can separate two classes of interfering intra-cycle quantum paths that evolve on two different time scales. In our experiment the two-dimensional character of the OTC field maps their contributions to different regions in momentum space facilitating their disentanglement. 
We envision that the coupling of sub-cycle time and space provided by OTC fields, as demonstrated by our experiment, can be exploited also for observing attosecond phenomena that lead to spatio-temporal variations of charge density. 
Careful analysis of intra-cycle interferences in OTC fields should thus permit the observation of sub-cycle variations of the ionization probability, caused, for example, by hole-wavepackets after ionization \cite{Loh2013,  Goulielmakis2010}. 
This might be of particular interest for studying charge density oscillations in molecules that are properly oriented with respect to the spatio-temporal evolution of the OTC field.
We believe that photoelectron intra-cycle interferometry has a potential for resolving attosecond dynamics that is largely untapped. OTC pulses hold a key to unlock this potential.



We acknowledge funding by the Austrian Science Fund (FWF) under grants P28475-N27, P25615-N27, P21463-N22, SFB-F49 NEXTlite, by the National Basic Research Program of China (grants 2013CB922201) and the NNSF of China (grant 11334009, 11474321). Financial support from the National Science and Engineering Research Council Discovery Grant No. 419092-2013-RGPIN is gratefully acknowledged.


\begin{thebibliography}{43}
\expandafter\ifx\csname natexlab\endcsname\relax\def\natexlab#1{#1}\fi
\expandafter\ifx\csname bibnamefont\endcsname\relax
  \def\bibnamefont#1{#1}\fi
\expandafter\ifx\csname bibfnamefont\endcsname\relax
  \def\bibfnamefont#1{#1}\fi
\expandafter\ifx\csname citenamefont\endcsname\relax
  \def\citenamefont#1{#1}\fi
\expandafter\ifx\csname url\endcsname\relax
  \def\url#1{\texttt{#1}}\fi
\expandafter\ifx\csname urlprefix\endcsname\relax\def\urlprefix{URL }\fi
\providecommand{\bibinfo}[2]{#2}
\providecommand{\eprint}[2][]{\url{#2}}

\bibitem[{\citenamefont{Kitzler and Lezius}(2005)}]{Kitzler2005}
\bibinfo{author}{\bibfnamefont{M.}~\bibnamefont{Kitzler}} \bibnamefont{and}
  \bibinfo{author}{\bibfnamefont{M.}~\bibnamefont{Lezius}},
  \bibinfo{journal}{Phys. Rev. Lett.} \textbf{\bibinfo{volume}{95}},
  \bibinfo{pages}{253001} (\bibinfo{year}{2005}), ISSN
  \bibinfo{issn}{0031-9007},
  \urlprefix\url{http://link.aps.org/doi/10.1103/PhysRevLett.95.253001}.

\bibitem[{\citenamefont{Kitzler et~al.}(2007)\citenamefont{Kitzler, Xie,
  Scrinzi, and Baltuska}}]{Kitzler2007}
\bibinfo{author}{\bibfnamefont{M.}~\bibnamefont{Kitzler}},
  \bibinfo{author}{\bibfnamefont{X.}~\bibnamefont{Xie}},
  \bibinfo{author}{\bibfnamefont{A.}~\bibnamefont{Scrinzi}}, \bibnamefont{and}
  \bibinfo{author}{\bibfnamefont{A.}~\bibnamefont{Baltuska}},
  \bibinfo{journal}{Phys. Rev. A} \textbf{\bibinfo{volume}{76}},
  \bibinfo{pages}{011801} (\bibinfo{year}{2007}), ISSN
  \bibinfo{issn}{1050-2947},
  \urlprefix\url{http://link.aps.org/doi/10.1103/PhysRevA.76.011801}.

\bibitem[{\citenamefont{Kitzler et~al.}(2008)\citenamefont{Kitzler, Xie,
  Roither, Scrinzi, and Baltuska}}]{Kitzler2008NJP}
\bibinfo{author}{\bibfnamefont{M.}~\bibnamefont{Kitzler}},
  \bibinfo{author}{\bibfnamefont{X.}~\bibnamefont{Xie}},
  \bibinfo{author}{\bibfnamefont{S.}~\bibnamefont{Roither}},
  \bibinfo{author}{\bibfnamefont{A.}~\bibnamefont{Scrinzi}}, \bibnamefont{and}
  \bibinfo{author}{\bibfnamefont{A.}~\bibnamefont{Baltuska}},
  \bibinfo{journal}{New Journal of Physics} \textbf{\bibinfo{volume}{10}},
  \bibinfo{pages}{025029} (\bibinfo{year}{2008}),
  \urlprefix\url{http://stacks.iop.org/1367-2630/10/i=2/a=025029}.

\bibitem[{\citenamefont{Comtois et~al.}(2013)\citenamefont{Comtois, Bandulet,
  Spanner, Pavi{\v{c}}i{\'{c}}, Meckel, Zeidler, P{\'{e}}pin, D{\"{o}}rner,
  Kieffer, Villeneuve et~al.}}]{Comtois2013}
\bibinfo{author}{\bibfnamefont{D.}~\bibnamefont{Comtois}},
  \bibinfo{author}{\bibfnamefont{H.-C.} \bibnamefont{Bandulet}},
  \bibinfo{author}{\bibfnamefont{M.}~\bibnamefont{Spanner}},
  \bibinfo{author}{\bibfnamefont{D.}~\bibnamefont{Pavi{\v{c}}i{\'{c}}}},
  \bibinfo{author}{\bibfnamefont{M.}~\bibnamefont{Meckel}},
  \bibinfo{author}{\bibfnamefont{D.}~\bibnamefont{Zeidler}},
  \bibinfo{author}{\bibfnamefont{H.}~\bibnamefont{P{\'{e}}pin}},
  \bibinfo{author}{\bibfnamefont{R.}~\bibnamefont{D{\"{o}}rner}},
  \bibinfo{author}{\bibfnamefont{J.-C.} \bibnamefont{Kieffer}},
  \bibinfo{author}{\bibfnamefont{D.}~\bibnamefont{Villeneuve}},
  \bibnamefont{et~al.}, \bibinfo{journal}{J. Mod. Opt.}
  \textbf{\bibinfo{volume}{60}}, \bibinfo{pages}{1395} (\bibinfo{year}{2013}),
  ISSN \bibinfo{issn}{0950-0340},
  \urlprefix\url{http://www.tandfonline.com/doi/abs/10.1080/09500340.2013.771755}.

\bibitem[{\citenamefont{Meckel et~al.}(2014)\citenamefont{Meckel, Staudte,
  Patchkovskii, Villeneuve, Corkum, and Spanner}}]{Meckel2014}
\bibinfo{author}{\bibfnamefont{M.}~\bibnamefont{Meckel}},
  \bibinfo{author}{\bibfnamefont{A.}~\bibnamefont{Staudte}},
  \bibinfo{author}{\bibfnamefont{S.}~\bibnamefont{Patchkovskii}},
  \bibinfo{author}{\bibfnamefont{D.~M.} \bibnamefont{Villeneuve}},
  \bibinfo{author}{\bibfnamefont{P.~B.} \bibnamefont{Corkum}},
  \bibnamefont{and} \bibinfo{author}{\bibfnamefont{M.}~\bibnamefont{Spanner}},
  \bibinfo{journal}{Nat. Phys.} \textbf{\bibinfo{volume}{10}},
  \bibinfo{pages}{594} (\bibinfo{year}{2014}), ISSN \bibinfo{issn}{1745-2473}.

\bibitem[{\citenamefont{Milo{\v{s}}evi{\'{c}}
  et~al.}(2006)\citenamefont{Milo{\v{s}}evi{\'{c}}, Paulus, Bauer, and
  Becker}}]{Milosevic2006}
\bibinfo{author}{\bibfnamefont{D.~B.} \bibnamefont{Milo{\v{s}}evi{\'{c}}}},
  \bibinfo{author}{\bibfnamefont{G.~G.} \bibnamefont{Paulus}},
  \bibinfo{author}{\bibfnamefont{D.}~\bibnamefont{Bauer}}, \bibnamefont{and}
  \bibinfo{author}{\bibfnamefont{W.}~\bibnamefont{Becker}},
  \bibinfo{journal}{J. Phys. B At. Mol. Opt. Phys.}
  \textbf{\bibinfo{volume}{39}}, \bibinfo{pages}{R203} (\bibinfo{year}{2006}),
  ISSN \bibinfo{issn}{0953-4075},
  \urlprefix\url{http://stacks.iop.org/0953-4075/39/i=14/a=R01
  http://stacks.iop.org/0953-4075/39/i=14/a=R01?key=crossref.212b8163f018982d1c4d95988bf902ce}.

\bibitem[{\citenamefont{Xie et~al.}(2012)\citenamefont{Xie, Roither, Kartashov,
  Persson, Arb{\'{o}}, Zhang, Gr{\"{a}}fe, Sch{\"{o}}ffler, Burgd{\"{o}}rfer,
  Baltu{\v{s}}ka et~al.}}]{Xie2012_interferometry}
\bibinfo{author}{\bibfnamefont{X.}~\bibnamefont{Xie}},
  \bibinfo{author}{\bibfnamefont{S.}~\bibnamefont{Roither}},
  \bibinfo{author}{\bibfnamefont{D.}~\bibnamefont{Kartashov}},
  \bibinfo{author}{\bibfnamefont{E.}~\bibnamefont{Persson}},
  \bibinfo{author}{\bibfnamefont{D.}~\bibnamefont{Arb{\'{o}}}},
  \bibinfo{author}{\bibfnamefont{L.}~\bibnamefont{Zhang}},
  \bibinfo{author}{\bibfnamefont{S.}~\bibnamefont{Gr{\"{a}}fe}},
  \bibinfo{author}{\bibfnamefont{M.}~\bibnamefont{Sch{\"{o}}ffler}},
  \bibinfo{author}{\bibfnamefont{J.}~\bibnamefont{Burgd{\"{o}}rfer}},
  \bibinfo{author}{\bibfnamefont{A.}~\bibnamefont{Baltu{\v{s}}ka}},
  \bibnamefont{et~al.}, \bibinfo{journal}{Phys. Rev. Lett.}
  \textbf{\bibinfo{volume}{108}}, \bibinfo{pages}{193004}
  (\bibinfo{year}{2012}), ISSN \bibinfo{issn}{0031-9007},
  \urlprefix\url{http://link.aps.org/doi/10.1103/PhysRevLett.108.193004}.

\bibitem[{\citenamefont{Arb{\'{o}} et~al.}(2010)\citenamefont{Arb{\'{o}},
  Ishikawa, Schiessl, Persson, and Burgd{\"{o}}rfer}}]{Arbo2010}
\bibinfo{author}{\bibfnamefont{D.~G.} \bibnamefont{Arb{\'{o}}}},
  \bibinfo{author}{\bibfnamefont{K.~L.} \bibnamefont{Ishikawa}},
  \bibinfo{author}{\bibfnamefont{K.}~\bibnamefont{Schiessl}},
  \bibinfo{author}{\bibfnamefont{E.}~\bibnamefont{Persson}}, \bibnamefont{and}
  \bibinfo{author}{\bibfnamefont{J.}~\bibnamefont{Burgd{\"{o}}rfer}},
  \bibinfo{journal}{Phys. Rev. A} \textbf{\bibinfo{volume}{81}},
  \bibinfo{pages}{021403} (\bibinfo{year}{2010}), ISSN
  \bibinfo{issn}{1050-2947},
  \urlprefix\url{http://link.aps.org/doi/10.1103/PhysRevA.81.021403}.

\bibitem[{\citenamefont{Huismans et~al.}(2011)\citenamefont{Huismans,
  Rouz{\'e}e, Gijsbertsen, Jungmann, Smolkowska, Logman, L{\'e}pine, Cauchy,
  Zamith, Marchenko et~al.}}]{Huismans2011Science}
\bibinfo{author}{\bibfnamefont{Y.}~\bibnamefont{Huismans}},
  \bibinfo{author}{\bibfnamefont{A.}~\bibnamefont{Rouz{\'e}e}},
  \bibinfo{author}{\bibfnamefont{A.}~\bibnamefont{Gijsbertsen}},
  \bibinfo{author}{\bibfnamefont{J.~H.} \bibnamefont{Jungmann}},
  \bibinfo{author}{\bibfnamefont{A.~S.} \bibnamefont{Smolkowska}},
  \bibinfo{author}{\bibfnamefont{P.~S. W.~M.} \bibnamefont{Logman}},
  \bibinfo{author}{\bibfnamefont{F.}~\bibnamefont{L{\'e}pine}},
  \bibinfo{author}{\bibfnamefont{C.}~\bibnamefont{Cauchy}},
  \bibinfo{author}{\bibfnamefont{S.}~\bibnamefont{Zamith}},
  \bibinfo{author}{\bibfnamefont{T.}~\bibnamefont{Marchenko}},
  \bibnamefont{et~al.}, \bibinfo{journal}{Science}
  \textbf{\bibinfo{volume}{331}}, \bibinfo{pages}{61} (\bibinfo{year}{2011}),
  ISSN \bibinfo{issn}{0036-8075},
  \urlprefix\url{http://science.sciencemag.org/content/331/6013/61}.

\bibitem[{\citenamefont{Haertelt et~al.}(2016)\citenamefont{Haertelt, Bian,
  Spanner, Staudte, and Corkum}}]{Haertelt2016PRL}
\bibinfo{author}{\bibfnamefont{M.}~\bibnamefont{Haertelt}},
  \bibinfo{author}{\bibfnamefont{X.-B.} \bibnamefont{Bian}},
  \bibinfo{author}{\bibfnamefont{M.}~\bibnamefont{Spanner}},
  \bibinfo{author}{\bibfnamefont{A.}~\bibnamefont{Staudte}}, \bibnamefont{and}
  \bibinfo{author}{\bibfnamefont{P.~B.} \bibnamefont{Corkum}},
  \bibinfo{journal}{Phys. Rev. Lett.} \textbf{\bibinfo{volume}{116}},
  \bibinfo{pages}{133001} (\bibinfo{year}{2016}),
  \urlprefix\url{https://link.aps.org/doi/10.1103/PhysRevLett.116.133001}.

\bibitem[{\citenamefont{Ray et~al.}(2008)\citenamefont{Ray, Ulrich, Bocharova,
  Maharjan, Ranitovic, Gramkow, Magrakvelidze, De, Litvinyuk, Le
  et~al.}}]{Ray2008PRL}
\bibinfo{author}{\bibfnamefont{D.}~\bibnamefont{Ray}},
  \bibinfo{author}{\bibfnamefont{B.}~\bibnamefont{Ulrich}},
  \bibinfo{author}{\bibfnamefont{I.}~\bibnamefont{Bocharova}},
  \bibinfo{author}{\bibfnamefont{C.}~\bibnamefont{Maharjan}},
  \bibinfo{author}{\bibfnamefont{P.}~\bibnamefont{Ranitovic}},
  \bibinfo{author}{\bibfnamefont{B.}~\bibnamefont{Gramkow}},
  \bibinfo{author}{\bibfnamefont{M.}~\bibnamefont{Magrakvelidze}},
  \bibinfo{author}{\bibfnamefont{S.}~\bibnamefont{De}},
  \bibinfo{author}{\bibfnamefont{I.~V.} \bibnamefont{Litvinyuk}},
  \bibinfo{author}{\bibfnamefont{A.~T.} \bibnamefont{Le}},
  \bibnamefont{et~al.}, \bibinfo{journal}{Phys. Rev. Lett.}
  \textbf{\bibinfo{volume}{100}}, \bibinfo{pages}{143002}
  (\bibinfo{year}{2008}),
  \urlprefix\url{https://link.aps.org/doi/10.1103/PhysRevLett.100.143002}.

\bibitem[{\citenamefont{Okunishi et~al.}(2008)\citenamefont{Okunishi,
  Morishita, Pr\"umper, Shimada, Lin, Watanabe, and Ueda}}]{Okunishi2008PRL}
\bibinfo{author}{\bibfnamefont{M.}~\bibnamefont{Okunishi}},
  \bibinfo{author}{\bibfnamefont{T.}~\bibnamefont{Morishita}},
  \bibinfo{author}{\bibfnamefont{G.}~\bibnamefont{Pr\"umper}},
  \bibinfo{author}{\bibfnamefont{K.}~\bibnamefont{Shimada}},
  \bibinfo{author}{\bibfnamefont{C.~D.} \bibnamefont{Lin}},
  \bibinfo{author}{\bibfnamefont{S.}~\bibnamefont{Watanabe}}, \bibnamefont{and}
  \bibinfo{author}{\bibfnamefont{K.}~\bibnamefont{Ueda}},
  \bibinfo{journal}{Phys. Rev. Lett.} \textbf{\bibinfo{volume}{100}},
  \bibinfo{pages}{143001} (\bibinfo{year}{2008}),
  \urlprefix\url{https://link.aps.org/doi/10.1103/PhysRevLett.100.143001}.

\bibitem[{\citenamefont{Meckel et~al.}(2008)\citenamefont{Meckel, Comtois,
  Zeidler, Staudte, Pavicic, Bandulet, P{\'{e}}pin, Kieffer, D{\"{o}}rner,
  Villeneuve et~al.}}]{Meckel2008}
\bibinfo{author}{\bibfnamefont{M.}~\bibnamefont{Meckel}},
  \bibinfo{author}{\bibfnamefont{D.}~\bibnamefont{Comtois}},
  \bibinfo{author}{\bibfnamefont{D.}~\bibnamefont{Zeidler}},
  \bibinfo{author}{\bibfnamefont{A.}~\bibnamefont{Staudte}},
  \bibinfo{author}{\bibfnamefont{D.}~\bibnamefont{Pavicic}},
  \bibinfo{author}{\bibfnamefont{H.~C.} \bibnamefont{Bandulet}},
  \bibinfo{author}{\bibfnamefont{H.}~\bibnamefont{P{\'{e}}pin}},
  \bibinfo{author}{\bibfnamefont{J.~C.} \bibnamefont{Kieffer}},
  \bibinfo{author}{\bibfnamefont{R.}~\bibnamefont{D{\"{o}}rner}},
  \bibinfo{author}{\bibfnamefont{D.~M.} \bibnamefont{Villeneuve}},
  \bibnamefont{et~al.}, \bibinfo{journal}{Science}
  \textbf{\bibinfo{volume}{320}}, \bibinfo{pages}{1478} (\bibinfo{year}{2008}),
  ISSN \bibinfo{issn}{1095-9203},
  \urlprefix\url{http://www.ncbi.nlm.nih.gov/pubmed/18556555
  http://www.sciencemag.org/content/320/5882/1478.abstract}.

\bibitem[{\citenamefont{Xu et~al.}(2010)\citenamefont{Xu, Chen, Le, and
  Lin}}]{Xu2010b}
\bibinfo{author}{\bibfnamefont{J.}~\bibnamefont{Xu}},
  \bibinfo{author}{\bibfnamefont{Z.}~\bibnamefont{Chen}},
  \bibinfo{author}{\bibfnamefont{A.-T.} \bibnamefont{Le}}, \bibnamefont{and}
  \bibinfo{author}{\bibfnamefont{C.~D.} \bibnamefont{Lin}},
  \bibinfo{journal}{Phys. Rev. A} \textbf{\bibinfo{volume}{82}},
  \bibinfo{pages}{033403} (\bibinfo{year}{2010}), ISSN
  \bibinfo{issn}{1050-2947},
  \urlprefix\url{http://link.aps.org/doi/10.1103/PhysRevA.82.033403
  https://link.aps.org/doi/10.1103/PhysRevA.82.033403}.

\bibitem[{\citenamefont{Blaga et~al.}(2012)\citenamefont{Blaga, Xu, DiChiara,
  Sistrunk, Zhang, Agostini, Miller, DiMauro, and Lin}}]{Blaga2012Nature}
\bibinfo{author}{\bibfnamefont{C.~I.} \bibnamefont{Blaga}},
  \bibinfo{author}{\bibfnamefont{J.}~\bibnamefont{Xu}},
  \bibinfo{author}{\bibfnamefont{A.~D.} \bibnamefont{DiChiara}},
  \bibinfo{author}{\bibfnamefont{E.}~\bibnamefont{Sistrunk}},
  \bibinfo{author}{\bibfnamefont{K.}~\bibnamefont{Zhang}},
  \bibinfo{author}{\bibfnamefont{P.}~\bibnamefont{Agostini}},
  \bibinfo{author}{\bibfnamefont{T.~A.} \bibnamefont{Miller}},
  \bibinfo{author}{\bibfnamefont{L.~F.} \bibnamefont{DiMauro}},
  \bibnamefont{and} \bibinfo{author}{\bibfnamefont{C.~D.} \bibnamefont{Lin}},
  \bibinfo{journal}{Nature} \textbf{\bibinfo{volume}{483}},
  \bibinfo{pages}{194} (\bibinfo{year}{2012}), ISSN \bibinfo{issn}{0028-0836},
  \urlprefix\url{http:https://dx.doi.org/10.1038/nature10820}.

\bibitem[{\citenamefont{Pullen et~al.}(2015)\citenamefont{Pullen, Wolter, Le,
  Baudisch, Hemmer, Senftleben, Schröter, Ullrich, Moshammer, Lin
  et~al.}}]{Pullen2015NatComm}
\bibinfo{author}{\bibfnamefont{M.~G.} \bibnamefont{Pullen}},
  \bibinfo{author}{\bibfnamefont{B.}~\bibnamefont{Wolter}},
  \bibinfo{author}{\bibfnamefont{A.-T.} \bibnamefont{Le}},
  \bibinfo{author}{\bibfnamefont{M.}~\bibnamefont{Baudisch}},
  \bibinfo{author}{\bibfnamefont{M.}~\bibnamefont{Hemmer}},
  \bibinfo{author}{\bibfnamefont{A.}~\bibnamefont{Senftleben}},
  \bibinfo{author}{\bibfnamefont{C.~D.} \bibnamefont{Schröter}},
  \bibinfo{author}{\bibfnamefont{J.}~\bibnamefont{Ullrich}},
  \bibinfo{author}{\bibfnamefont{R.}~\bibnamefont{Moshammer}},
  \bibinfo{author}{\bibfnamefont{C.~D.} \bibnamefont{Lin}},
  \bibnamefont{et~al.}, \bibinfo{journal}{Nature Communications}
  \textbf{\bibinfo{volume}{6}}, \bibinfo{pages}{7262} (\bibinfo{year}{2015}),
  ISSN \bibinfo{issn}{2041-1733},
  \urlprefix\url{http:https://dx.doi.org/10.1038/ncomms8262}.

\bibitem[{\citenamefont{Wolter et~al.}(2016)\citenamefont{Wolter, Pullen, Le,
  Baudisch, Doblhoff-Dier, Senftleben, Hemmer, Schroter, Ullrich, Pfeifer
  et~al.}}]{Wolter2016}
\bibinfo{author}{\bibfnamefont{B.}~\bibnamefont{Wolter}},
  \bibinfo{author}{\bibfnamefont{M.~G.} \bibnamefont{Pullen}},
  \bibinfo{author}{\bibfnamefont{A.-T.} \bibnamefont{Le}},
  \bibinfo{author}{\bibfnamefont{M.}~\bibnamefont{Baudisch}},
  \bibinfo{author}{\bibfnamefont{K.}~\bibnamefont{Doblhoff-Dier}},
  \bibinfo{author}{\bibfnamefont{A.}~\bibnamefont{Senftleben}},
  \bibinfo{author}{\bibfnamefont{M.}~\bibnamefont{Hemmer}},
  \bibinfo{author}{\bibfnamefont{C.~D.} \bibnamefont{Schroter}},
  \bibinfo{author}{\bibfnamefont{J.}~\bibnamefont{Ullrich}},
  \bibinfo{author}{\bibfnamefont{T.}~\bibnamefont{Pfeifer}},
  \bibnamefont{et~al.}, \bibinfo{journal}{Science (80-. ).}
  \textbf{\bibinfo{volume}{354}}, \bibinfo{pages}{308} (\bibinfo{year}{2016}),
  ISSN \bibinfo{issn}{0036-8075},
  \urlprefix\url{http://www.sciencemag.org/cgi/doi/10.1126/science.aah3429}.

\bibitem[{\citenamefont{Agostini et~al.}(1979)\citenamefont{Agostini, Fabre,
  Mainfray, Petite, and Rahman}}]{Agostini1979}
\bibinfo{author}{\bibfnamefont{P.}~\bibnamefont{Agostini}},
  \bibinfo{author}{\bibfnamefont{F.}~\bibnamefont{Fabre}},
  \bibinfo{author}{\bibfnamefont{G.}~\bibnamefont{Mainfray}},
  \bibinfo{author}{\bibfnamefont{G.}~\bibnamefont{Petite}}, \bibnamefont{and}
  \bibinfo{author}{\bibfnamefont{N.}~\bibnamefont{Rahman}},
  \bibinfo{journal}{Phys. Rev. Lett.} \textbf{\bibinfo{volume}{42}},
  \bibinfo{pages}{1127} (\bibinfo{year}{1979}), ISSN \bibinfo{issn}{0031-9007},
  \urlprefix\url{http://link.aps.org/doi/10.1103/PhysRevLett.42.1127}.

\bibitem[{\citenamefont{Freeman and Bucksbaum}(1991)}]{Freeman1991}
\bibinfo{author}{\bibfnamefont{R.~R.} \bibnamefont{Freeman}} \bibnamefont{and}
  \bibinfo{author}{\bibfnamefont{P.~H.} \bibnamefont{Bucksbaum}},
  \bibinfo{journal}{J. Phys. B At. Mol. Opt. Phys.}
  \textbf{\bibinfo{volume}{24}}, \bibinfo{pages}{325} (\bibinfo{year}{1991}),
  ISSN \bibinfo{issn}{0953-4075},
  \urlprefix\url{http://stacks.iop.org/0953-4075/24/i=2/a=004?key=crossref.3e52b19acc1d33cd17daa8b5e88c99c6}.

\bibitem[{\citenamefont{Arb{\'{o}} et~al.}(2006)\citenamefont{Arb{\'{o}},
  Persson, and Burgd{\"{o}}rfer}}]{Arbo2006a}
\bibinfo{author}{\bibfnamefont{D.~G.} \bibnamefont{Arb{\'{o}}}},
  \bibinfo{author}{\bibfnamefont{E.}~\bibnamefont{Persson}}, \bibnamefont{and}
  \bibinfo{author}{\bibfnamefont{J.}~\bibnamefont{Burgd{\"{o}}rfer}},
  \bibinfo{journal}{Phys. Rev. A} \textbf{\bibinfo{volume}{74}},
  \bibinfo{pages}{063407} (\bibinfo{year}{2006}), ISSN
  \bibinfo{issn}{1050-2947},
  \urlprefix\url{http://link.aps.org/doi/10.1103/PhysRevA.74.063407}.

\bibitem[{\citenamefont{Arb{\'{o}} et~al.}(2012)\citenamefont{Arb{\'{o}},
  Ishikawa, Persson, and Burgd{\"{o}}rfer}}]{Arbo2012}
\bibinfo{author}{\bibfnamefont{D.~G.} \bibnamefont{Arb{\'{o}}}},
  \bibinfo{author}{\bibfnamefont{K.~L.} \bibnamefont{Ishikawa}},
  \bibinfo{author}{\bibfnamefont{E.}~\bibnamefont{Persson}}, \bibnamefont{and}
  \bibinfo{author}{\bibfnamefont{J.}~\bibnamefont{Burgd{\"{o}}rfer}},
  \bibinfo{journal}{Nucl. Instruments Methods Phys. Res. Sect. B Beam Interact.
  with Mater. Atoms} \textbf{\bibinfo{volume}{279}}, \bibinfo{pages}{24}
  (\bibinfo{year}{2012}), ISSN \bibinfo{issn}{0168583X}, \eprint{0912.5470},
  \urlprefix\url{http://linkinghub.elsevier.com/retrieve/pii/S0168583X11009694}.

\bibitem[{\citenamefont{Lindner et~al.}(2005)\citenamefont{Lindner, Sch\"atzel,
  Walther, Baltu\ifmmode~\check{s}\else \v{s}\fi{}ka, Goulielmakis, Krausz,
  Milo\ifmmode \check{s}\else \v{s}\fi{}evi\ifmmode~\acute{c}\else \'{c}\fi{},
  Bauer, Becker, and Paulus}}]{Lindner2005PRL}
\bibinfo{author}{\bibfnamefont{F.}~\bibnamefont{Lindner}},
  \bibinfo{author}{\bibfnamefont{M.~G.} \bibnamefont{Sch\"atzel}},
  \bibinfo{author}{\bibfnamefont{H.}~\bibnamefont{Walther}},
  \bibinfo{author}{\bibfnamefont{A.}~\bibnamefont{Baltu\ifmmode~\check{s}\else
  \v{s}\fi{}ka}},
  \bibinfo{author}{\bibfnamefont{E.}~\bibnamefont{Goulielmakis}},
  \bibinfo{author}{\bibfnamefont{F.}~\bibnamefont{Krausz}},
  \bibinfo{author}{\bibfnamefont{D.~B.} \bibnamefont{Milo\ifmmode
  \check{s}\else \v{s}\fi{}evi\ifmmode~\acute{c}\else \'{c}\fi{}}},
  \bibinfo{author}{\bibfnamefont{D.}~\bibnamefont{Bauer}},
  \bibinfo{author}{\bibfnamefont{W.}~\bibnamefont{Becker}}, \bibnamefont{and}
  \bibinfo{author}{\bibfnamefont{G.~G.} \bibnamefont{Paulus}},
  \bibinfo{journal}{Phys. Rev. Lett.} \textbf{\bibinfo{volume}{95}},
  \bibinfo{pages}{040401} (\bibinfo{year}{2005}),
  \urlprefix\url{https://link.aps.org/doi/10.1103/PhysRevLett.95.040401}.

\bibitem[{\citenamefont{Gopal et~al.}(2009)\citenamefont{Gopal, Simeonidis,
  Moshammer, Ergler, D{\"{u}}rr, Kurka, K{\"{u}}hnel, Tschuch, Schr{\"{o}}ter,
  Bauer et~al.}}]{Gopal2009}
\bibinfo{author}{\bibfnamefont{R.}~\bibnamefont{Gopal}},
  \bibinfo{author}{\bibfnamefont{K.}~\bibnamefont{Simeonidis}},
  \bibinfo{author}{\bibfnamefont{R.}~\bibnamefont{Moshammer}},
  \bibinfo{author}{\bibfnamefont{T.}~\bibnamefont{Ergler}},
  \bibinfo{author}{\bibfnamefont{M.}~\bibnamefont{D{\"{u}}rr}},
  \bibinfo{author}{\bibfnamefont{M.}~\bibnamefont{Kurka}},
  \bibinfo{author}{\bibfnamefont{K.-U.} \bibnamefont{K{\"{u}}hnel}},
  \bibinfo{author}{\bibfnamefont{S.}~\bibnamefont{Tschuch}},
  \bibinfo{author}{\bibfnamefont{C.-D.} \bibnamefont{Schr{\"{o}}ter}},
  \bibinfo{author}{\bibfnamefont{D.}~\bibnamefont{Bauer}},
  \bibnamefont{et~al.}, \bibinfo{journal}{Phys. Rev. Lett.}
  \textbf{\bibinfo{volume}{103}}, \bibinfo{pages}{053001}
  (\bibinfo{year}{2009}), ISSN \bibinfo{issn}{0031-9007},
  \urlprefix\url{http://link.aps.org/doi/10.1103/PhysRevLett.103.053001}.

\bibitem[{\citenamefont{Zhang et~al.}(2014)\citenamefont{Zhang, Xie, Roither,
  Kartashov, Wang, Wang, Sch\"offler, Shafir, Corkum,
  Baltu\ifmmode~\check{s}\else \v{s}\fi{}ka et~al.}}]{Zhang2014PRA}
\bibinfo{author}{\bibfnamefont{L.}~\bibnamefont{Zhang}},
  \bibinfo{author}{\bibfnamefont{X.}~\bibnamefont{Xie}},
  \bibinfo{author}{\bibfnamefont{S.}~\bibnamefont{Roither}},
  \bibinfo{author}{\bibfnamefont{D.}~\bibnamefont{Kartashov}},
  \bibinfo{author}{\bibfnamefont{Y.}~\bibnamefont{Wang}},
  \bibinfo{author}{\bibfnamefont{C.}~\bibnamefont{Wang}},
  \bibinfo{author}{\bibfnamefont{M.}~\bibnamefont{Sch\"offler}},
  \bibinfo{author}{\bibfnamefont{D.}~\bibnamefont{Shafir}},
  \bibinfo{author}{\bibfnamefont{P.~B.} \bibnamefont{Corkum}},
  \bibinfo{author}{\bibfnamefont{A.}~\bibnamefont{Baltu\ifmmode~\check{s}\else
  \v{s}\fi{}ka}}, \bibnamefont{et~al.}, \bibinfo{journal}{Phys. Rev. A}
  \textbf{\bibinfo{volume}{90}}, \bibinfo{pages}{061401}
  (\bibinfo{year}{2014}),
  \urlprefix\url{https://link.aps.org/doi/10.1103/PhysRevA.90.061401}.

\bibitem[{\citenamefont{Richter et~al.}(2015)\citenamefont{Richter, Kunitski,
  Sch\"offler, Jahnke, Schmidt, Li, Liu, and D\"orner}}]{Richter2015PRL}
\bibinfo{author}{\bibfnamefont{M.}~\bibnamefont{Richter}},
  \bibinfo{author}{\bibfnamefont{M.}~\bibnamefont{Kunitski}},
  \bibinfo{author}{\bibfnamefont{M.}~\bibnamefont{Sch\"offler}},
  \bibinfo{author}{\bibfnamefont{T.}~\bibnamefont{Jahnke}},
  \bibinfo{author}{\bibfnamefont{L.~P.~H.} \bibnamefont{Schmidt}},
  \bibinfo{author}{\bibfnamefont{M.}~\bibnamefont{Li}},
  \bibinfo{author}{\bibfnamefont{Y.}~\bibnamefont{Liu}}, \bibnamefont{and}
  \bibinfo{author}{\bibfnamefont{R.}~\bibnamefont{D\"orner}},
  \bibinfo{journal}{Phys. Rev. Lett.} \textbf{\bibinfo{volume}{114}},
  \bibinfo{pages}{143001} (\bibinfo{year}{2015}),
  \urlprefix\url{https://link.aps.org/doi/10.1103/PhysRevLett.114.143001}.

\bibitem[{\citenamefont{Geng et~al.}(2015)\citenamefont{Geng, Xiong, Xiao,
  Peng, and Gong}}]{Geng2015PRL}
\bibinfo{author}{\bibfnamefont{J.-W.} \bibnamefont{Geng}},
  \bibinfo{author}{\bibfnamefont{W.-H.} \bibnamefont{Xiong}},
  \bibinfo{author}{\bibfnamefont{X.-R.} \bibnamefont{Xiao}},
  \bibinfo{author}{\bibfnamefont{L.-Y.} \bibnamefont{Peng}}, \bibnamefont{and}
  \bibinfo{author}{\bibfnamefont{Q.}~\bibnamefont{Gong}},
  \bibinfo{journal}{Phys. Rev. Lett.} \textbf{\bibinfo{volume}{115}},
  \bibinfo{pages}{193001} (\bibinfo{year}{2015}),
  \urlprefix\url{https://link.aps.org/doi/10.1103/PhysRevLett.115.193001}.

\bibitem[{\citenamefont{Richter et~al.}(2016)\citenamefont{Richter, Kunitski,
  Sch\"offler, Jahnke, Schmidt, and D\"orner}}]{Richter2016PRA}
\bibinfo{author}{\bibfnamefont{M.}~\bibnamefont{Richter}},
  \bibinfo{author}{\bibfnamefont{M.}~\bibnamefont{Kunitski}},
  \bibinfo{author}{\bibfnamefont{M.}~\bibnamefont{Sch\"offler}},
  \bibinfo{author}{\bibfnamefont{T.}~\bibnamefont{Jahnke}},
  \bibinfo{author}{\bibfnamefont{L.~P.~H.} \bibnamefont{Schmidt}},
  \bibnamefont{and} \bibinfo{author}{\bibfnamefont{R.}~\bibnamefont{D\"orner}},
  \bibinfo{journal}{Phys. Rev. A} \textbf{\bibinfo{volume}{94}},
  \bibinfo{pages}{033416} (\bibinfo{year}{2016}),
  \urlprefix\url{https://link.aps.org/doi/10.1103/PhysRevA.94.033416}.

\bibitem[{\citenamefont{Li et~al.}(2015{\natexlab{a}})\citenamefont{Li, Geng,
  Liu, Zheng, Peng, Gong, and Liu}}]{Li2015PRA}
\bibinfo{author}{\bibfnamefont{M.}~\bibnamefont{Li}},
  \bibinfo{author}{\bibfnamefont{J.-W.} \bibnamefont{Geng}},
  \bibinfo{author}{\bibfnamefont{M.-M.} \bibnamefont{Liu}},
  \bibinfo{author}{\bibfnamefont{X.}~\bibnamefont{Zheng}},
  \bibinfo{author}{\bibfnamefont{L.-Y.} \bibnamefont{Peng}},
  \bibinfo{author}{\bibfnamefont{Q.}~\bibnamefont{Gong}}, \bibnamefont{and}
  \bibinfo{author}{\bibfnamefont{Y.}~\bibnamefont{Liu}},
  \bibinfo{journal}{Phys. Rev. A} \textbf{\bibinfo{volume}{92}},
  \bibinfo{pages}{013416} (\bibinfo{year}{2015}{\natexlab{a}}),
  \urlprefix\url{https://link.aps.org/doi/10.1103/PhysRevA.92.013416}.

\bibitem[{\citenamefont{D\"orner et~al.}(2000)\citenamefont{D\"orner, Mergel,
  Jagutzki, Spielberger, Ullrich, Moshammer, and
  Schmidt-B\"ocking}}]{Doerner2000PhysRep}
\bibinfo{author}{\bibfnamefont{R.}~\bibnamefont{D\"orner}},
  \bibinfo{author}{\bibfnamefont{V.}~\bibnamefont{Mergel}},
  \bibinfo{author}{\bibfnamefont{O.}~\bibnamefont{Jagutzki}},
  \bibinfo{author}{\bibfnamefont{L.}~\bibnamefont{Spielberger}},
  \bibinfo{author}{\bibfnamefont{J.}~\bibnamefont{Ullrich}},
  \bibinfo{author}{\bibfnamefont{R.}~\bibnamefont{Moshammer}},
  \bibnamefont{and}
  \bibinfo{author}{\bibfnamefont{H.}~\bibnamefont{Schmidt-B\"ocking}},
  \bibinfo{journal}{Physics Reports} \textbf{\bibinfo{volume}{330}},
  \bibinfo{pages}{95 } (\bibinfo{year}{2000}), ISSN \bibinfo{issn}{0370-1573},
  \urlprefix\url{http://www.sciencedirect.com/science/article/pii/S037015739900109X}.

\bibitem[{\citenamefont{{Van Linden van den Heuvell} and
  Muller}(1988)}]{VanLindenvandenHeuvell1988}
\bibinfo{author}{\bibfnamefont{H.}~\bibnamefont{{Van Linden van den Heuvell}}}
  \bibnamefont{and} \bibinfo{author}{\bibfnamefont{H.}~\bibnamefont{Muller}},
  in \emph{\bibinfo{booktitle}{Multiphot. Process. Proc. 4th Int. Conf.
  Multiphot. Process. JILA, Boulder, Color. July 13-17, 1987}}, edited by
  \bibinfo{editor}{\bibfnamefont{S.~J.} \bibnamefont{Smith}} \bibnamefont{and}
  \bibinfo{editor}{\bibfnamefont{P.~L.} \bibnamefont{Knight}}
  (\bibinfo{publisher}{Cambridge University Press}, \bibinfo{year}{1988}), pp.
  \bibinfo{pages}{25--34}.

\bibitem[{\citenamefont{Corkum et~al.}(1989)\citenamefont{Corkum, Burnett, and
  Brunel}}]{Corkum1989}
\bibinfo{author}{\bibfnamefont{P.}~\bibnamefont{Corkum}},
  \bibinfo{author}{\bibfnamefont{N.}~\bibnamefont{Burnett}}, \bibnamefont{and}
  \bibinfo{author}{\bibfnamefont{F.}~\bibnamefont{Brunel}},
  \bibinfo{journal}{Phys. Rev. Lett.} \textbf{\bibinfo{volume}{62}},
  \bibinfo{pages}{1259} (\bibinfo{year}{1989}), ISSN \bibinfo{issn}{0031-9007},
  \urlprefix\url{http://link.aps.org/doi/10.1103/PhysRevLett.62.1259}.

\bibitem[{\citenamefont{Reiss}(1980)}]{Reiss1980PRA}
\bibinfo{author}{\bibfnamefont{H.~R.} \bibnamefont{Reiss}},
  \bibinfo{journal}{Phys. Rev. A} \textbf{\bibinfo{volume}{22}},
  \bibinfo{pages}{1786} (\bibinfo{year}{1980}),
  \urlprefix\url{https://link.aps.org/doi/10.1103/PhysRevA.22.1786}.

\bibitem[{\citenamefont{Milo\ifmmode \check{s}\else
  \v{s}\fi{}evi\ifmmode~\acute{c}\else \'{c}\fi{} and
  Ehlotzky}(1998)}]{Milosevic1998PRA}
\bibinfo{author}{\bibfnamefont{D.~B.} \bibnamefont{Milo\ifmmode \check{s}\else
  \v{s}\fi{}evi\ifmmode~\acute{c}\else \'{c}\fi{}}} \bibnamefont{and}
  \bibinfo{author}{\bibfnamefont{F.}~\bibnamefont{Ehlotzky}},
  \bibinfo{journal}{Phys. Rev. A} \textbf{\bibinfo{volume}{58}},
  \bibinfo{pages}{3124} (\bibinfo{year}{1998}),
  \urlprefix\url{https://link.aps.org/doi/10.1103/PhysRevA.58.3124}.

\bibitem[{\citenamefont{de~Morrison~Faria
  et~al.}(2008)\citenamefont{de~Morrison~Faria, Shaaran, Liu, and
  Yang}}]{Figueira2008PRA}
\bibinfo{author}{\bibfnamefont{C.~F.} \bibnamefont{de~Morrison~Faria}},
  \bibinfo{author}{\bibfnamefont{T.}~\bibnamefont{Shaaran}},
  \bibinfo{author}{\bibfnamefont{X.}~\bibnamefont{Liu}}, \bibnamefont{and}
  \bibinfo{author}{\bibfnamefont{W.}~\bibnamefont{Yang}},
  \bibinfo{journal}{Phys. Rev. A} \textbf{\bibinfo{volume}{78}},
  \bibinfo{pages}{043407} (\bibinfo{year}{2008}).

\bibitem[{\citenamefont{Becker et~al.}(2002)\citenamefont{Becker, Grasbon,
  Kopold, Milo\v{s}evi\'c, Paulus, and Walther}}]{Becker2002AdvAtMolOptPhys}
\bibinfo{author}{\bibfnamefont{W.}~\bibnamefont{Becker}},
  \bibinfo{author}{\bibfnamefont{F.}~\bibnamefont{Grasbon}},
  \bibinfo{author}{\bibfnamefont{R.}~\bibnamefont{Kopold}},
  \bibinfo{author}{\bibfnamefont{D.~B.} \bibnamefont{Milo\v{s}evi\'c}},
  \bibinfo{author}{\bibfnamefont{G.~G.} \bibnamefont{Paulus}},
  \bibnamefont{and} \bibinfo{author}{\bibfnamefont{H.}~\bibnamefont{Walther}},
  \bibinfo{journal}{Adv. At. Mol. Opt. Phys.} \textbf{\bibinfo{volume}{48}},
  \bibinfo{pages}{35} (\bibinfo{year}{2002}),
  \urlprefix\url{https://doi.org/10.1016/S1049-250X(02)80006-4}.

\bibitem[{\citenamefont{Figueira~de Morisson~Faria
  et~al.}(2002)\citenamefont{Figueira~de Morisson~Faria, Schomerus, and
  Becker}}]{Figueira2002PRA}
\bibinfo{author}{\bibfnamefont{C.}~\bibnamefont{Figueira~de Morisson~Faria}},
  \bibinfo{author}{\bibfnamefont{H.}~\bibnamefont{Schomerus}},
  \bibnamefont{and} \bibinfo{author}{\bibfnamefont{W.}~\bibnamefont{Becker}},
  \bibinfo{journal}{Phys. Rev. A} \textbf{\bibinfo{volume}{66}},
  \bibinfo{pages}{043413} (\bibinfo{year}{2002}),
  \urlprefix\url{https://link.aps.org/doi/10.1103/PhysRevA.66.043413}.

\bibitem[{\citenamefont{Paulus et~al.}(2001)\citenamefont{Paulus, Grasbon,
  Walther, Kopold, and Becker}}]{Paulus2001a}
\bibinfo{author}{\bibfnamefont{G.~G.} \bibnamefont{Paulus}},
  \bibinfo{author}{\bibfnamefont{F.}~\bibnamefont{Grasbon}},
  \bibinfo{author}{\bibfnamefont{H.}~\bibnamefont{Walther}},
  \bibinfo{author}{\bibfnamefont{R.}~\bibnamefont{Kopold}}, \bibnamefont{and}
  \bibinfo{author}{\bibfnamefont{W.}~\bibnamefont{Becker}},
  \bibinfo{journal}{Phys. Rev. A} \textbf{\bibinfo{volume}{64}},
  \bibinfo{pages}{021401} (\bibinfo{year}{2001}), ISSN
  \bibinfo{issn}{1050-2947},
  \urlprefix\url{https://link.aps.org/doi/10.1103/PhysRevA.64.021401}.

\bibitem[{\citenamefont{Yu et~al.}(2016)\citenamefont{Yu, Wang, Lai, Huang,
  Quan, and Liu}}]{Yu2016a}
\bibinfo{author}{\bibfnamefont{S.}~\bibnamefont{Yu}},
  \bibinfo{author}{\bibfnamefont{Y.}~\bibnamefont{Wang}},
  \bibinfo{author}{\bibfnamefont{X.}~\bibnamefont{Lai}},
  \bibinfo{author}{\bibfnamefont{Y.}~\bibnamefont{Huang}},
  \bibinfo{author}{\bibfnamefont{W.}~\bibnamefont{Quan}}, \bibnamefont{and}
  \bibinfo{author}{\bibfnamefont{X.}~\bibnamefont{Liu}},
  \bibinfo{journal}{Phys. Rev. A} \textbf{\bibinfo{volume}{94}},
  \bibinfo{pages}{033418} (\bibinfo{year}{2016}), ISSN
  \bibinfo{issn}{2469-9926},
  \urlprefix\url{https://link.aps.org/doi/10.1103/PhysRevA.94.033418}.

\bibitem[{\citenamefont{Lai and Faria}(2013)}]{Lai2013PRA}
\bibinfo{author}{\bibfnamefont{X.}~\bibnamefont{Lai}} \bibnamefont{and}
  \bibinfo{author}{\bibfnamefont{C.~F. d.~M.} \bibnamefont{Faria}},
  \bibinfo{journal}{Phys. Rev. A} \textbf{\bibinfo{volume}{88}},
  \bibinfo{pages}{013406} (\bibinfo{year}{2013}),
  \urlprefix\url{https://link.aps.org/doi/10.1103/PhysRevA.88.013406}.

\bibitem[{\citenamefont{Li et~al.}(2015{\natexlab{b}})\citenamefont{Li, Sun,
  Xie, Shao, Deng, Wu, Gong, and Liu}}]{Li2015SR}
\bibinfo{author}{\bibfnamefont{M.}~\bibnamefont{Li}},
  \bibinfo{author}{\bibfnamefont{X.}~\bibnamefont{Sun}},
  \bibinfo{author}{\bibfnamefont{X.}~\bibnamefont{Xie}},
  \bibinfo{author}{\bibfnamefont{Y.}~\bibnamefont{Shao}},
  \bibinfo{author}{\bibfnamefont{Y.}~\bibnamefont{Deng}},
  \bibinfo{author}{\bibfnamefont{C.}~\bibnamefont{Wu}},
  \bibinfo{author}{\bibfnamefont{Q.}~\bibnamefont{Gong}}, \bibnamefont{and}
  \bibinfo{author}{\bibfnamefont{Y.}~\bibnamefont{Liu}},
  \bibinfo{journal}{Scientific Reports} \textbf{\bibinfo{volume}{5}},
  \bibinfo{pages}{8519} (\bibinfo{year}{2015}{\natexlab{b}}),
  \urlprefix\url{http:https://doi.org/10.1038/srep08519}.

\bibitem[{\citenamefont{Maxwell et~al.}(2017)\citenamefont{Maxwell,
  Al-Jawahiry, Das, and Faria}}]{Maxwell2017}
\bibinfo{author}{\bibfnamefont{A.~S.} \bibnamefont{Maxwell}},
  \bibinfo{author}{\bibfnamefont{A.}~\bibnamefont{Al-Jawahiry}},
  \bibinfo{author}{\bibfnamefont{T.}~\bibnamefont{Das}}, \bibnamefont{and}
  \bibinfo{author}{\bibfnamefont{C.~F. d.~M.} \bibnamefont{Faria}}, pp.
  \bibinfo{pages}{1--17} (\bibinfo{year}{2017}), \eprint{1705.01518},
  \urlprefix\url{http://arxiv.org/abs/1705.01518}.

\bibitem[{\citenamefont{Loh and Leone}(2013)}]{Loh2013}
\bibinfo{author}{\bibfnamefont{Z.-h.} \bibnamefont{Loh}} \bibnamefont{and}
  \bibinfo{author}{\bibfnamefont{S.~R.} \bibnamefont{Leone}},
  \bibinfo{journal}{J. Phys. Chem. Lett.} \textbf{\bibinfo{volume}{4}},
  \bibinfo{pages}{292} (\bibinfo{year}{2013}), ISSN \bibinfo{issn}{1948-7185},
  \urlprefix\url{http://pubs.acs.org/doi/abs/10.1021/jz301910n}.

\bibitem[{\citenamefont{Goulielmakis et~al.}(2010)\citenamefont{Goulielmakis,
  Loh, Wirth, Santra, Rohringer, Yakovlev, Zherebtsov, Pfeifer, Azzeer, Kling
  et~al.}}]{Goulielmakis2010}
\bibinfo{author}{\bibfnamefont{E.}~\bibnamefont{Goulielmakis}},
  \bibinfo{author}{\bibfnamefont{Z.-H.} \bibnamefont{Loh}},
  \bibinfo{author}{\bibfnamefont{A.}~\bibnamefont{Wirth}},
  \bibinfo{author}{\bibfnamefont{R.}~\bibnamefont{Santra}},
  \bibinfo{author}{\bibfnamefont{N.}~\bibnamefont{Rohringer}},
  \bibinfo{author}{\bibfnamefont{V.~S.} \bibnamefont{Yakovlev}},
  \bibinfo{author}{\bibfnamefont{S.}~\bibnamefont{Zherebtsov}},
  \bibinfo{author}{\bibfnamefont{T.}~\bibnamefont{Pfeifer}},
  \bibinfo{author}{\bibfnamefont{A.~M.} \bibnamefont{Azzeer}},
  \bibinfo{author}{\bibfnamefont{M.~F.} \bibnamefont{Kling}},
  \bibnamefont{et~al.}, \bibinfo{journal}{Nature}
  \textbf{\bibinfo{volume}{466}}, \bibinfo{pages}{739} (\bibinfo{year}{2010}),
  ISSN \bibinfo{issn}{0028-0836},
  \urlprefix\url{http://www.nature.com/doifinder/10.1038/nature09212}.

\end{thebibliography}

\end{document}